# Fault Detection Using Nonlinear Low-Dimensional Representation of Sensor Data

Kai Shen, SAS Institute Inc.

Anya Mcguirk, SAS Institute Inc.

Yuwei Liao, SAS Institute Inc.

Arin Chaudhuri, SAS Institute Inc.

Deovrat Kakde, SAS Institute Inc.

## 1 SUMMARY & CONCLUSIONS

Sensor data analysis plays a key role in health assessment of critical equipment. Such data are multivariate and exhibit nonlinear relationships. This paper describes how one can exploit nonlinear dimension reduction techniques, such as the t-distributed stochastic neighbor embedding (t-SNE) and kernel principal component analysis (KPCA) for fault detection. We show that using anomaly detection with low dimensional representations provides better interpretability and is conducive to edge processing in IoT applications.

## 2 INTRODUCTION

Real-time monitoring of equipment health using sensors for improving reliability and uptime is becoming prevalent in different industries. Recent advances in many enabling technologies such as sensing, computing and communication are instrumental in achieving this objective. Real-time health monitoring enables transitioning from traditional fixed schedule preventive maintenance to predictive maintenance, where decisions regarding maintenance are based on an objective assessment of the equipment health.

The reduction in price of sensors has enabled widespread adoption of sensor technology for health monitoring. In 2004 the average cost of sensors was $1.30 and in the year 2020, it is expected to come down to $0.38 [1]. Industries such as mining, transportation, and aerospace are among the leaders in adoption of sensor-enabled predictive maintenance. Equipment such as locomotives, airplanes, and earth moving equipment used in these industries tend to be capital intensive with high uptime requirement, hence it is essential to proactively monitor their health for impending faults or failures. This equipment is typically fitted with multiple sensors, with each sensor measuring a key health parameter such as temperature, pressure, or acceleration. The frequency of measurement depends on the health parameter, its expected fluctuation and resulting criticality. Certain parameters such as acceleration related to vibration, which are analyzed in the frequency domain, need to be measured more frequently, where a few thousand observations per second is not uncommon. To give an example, currently a typical airplane has about 7000 sensors for measuring important health parameters and these sensors creates 2.5 terabytes of data per day. By 2020, this number is expected to triple or quadruple to over 7.5 terabytes [2].

Often the relevant information provided by the many different sensors on these valuable pieces of equipment is highly correlated and sometimes even redundant. Principal components analysis (PCA) tends to be the preferred procedure to reduce the dimensionality of the problem. PCA assumes that the sensors are linearly related and that this linear relationship is similar in all operating modes of the equipment. In practice there is no reason to believe that either of these assumptions is reasonable in the context of sensor data for fault detection and predictive maintenance. The various operating modes of a machine can possibly be accounted for by first clustering and then using PCA, but oftentimes the number of effective operating modes is unknown. Further, the linearity assumption is still questionable. For example, sensor data collected from the chemical process industry often show a nonlinear relationship [3]. The nonlinear relationships between various sensor measurements are described in [3-4], which propose techniques such as kernel principal component analysis (KPCA) for analysis.

Thus sensor data can be characterized as multivariate and nonlinear, representing multiple operating modes, which are collected at a high frequency. Analytical techniques used for fault detection take sensor data as an input and produce probability of fault over a time line of interest. The probability provides actionable insights for intervention. We argue that visualizing sensor data together with fault prediction probability can help an analyst gain better insights regarding the equipment. As sensors are so numerous and the data multivariate and correlated, one may not get many insights by visualizing each sensor measurement individually. We propose two analytical techniques: t-distributed stochastic neighborhood embedding (t-SNE) and kernel principal component analysis (KPCA) for visualizing the multivariate sensor data. Both techniques can be used to visualize high-dimensional data in 2- or 3- dimensional space. We have observed that such low dimensional embedding yields important insights from the data. The rest of the paper is organized as follows. Section 3 introduces t-SNE and KPCA techniques. Section 4 describes sensor data sets and provides low-dimensional visualization using t-SNE and KPCA. Conclusions are provided in section 5.

## 3 NONLINEAR DIMENSION REDUCTION TECHNIQUES

t-SNE:

The t-distributed stochastic neighborhood embedding (t-SNE) is a dimension reduction technique useful for visualizing high dimensional data in a 2- or 3- dimensional feature map. This technique was introduced by van der Matten and Hinton in 2008[5] and is capable of capturing much of the local structure in the high dimensional data while revealing global structure as well. For example, if the data have multiple clusters, t-SNE can reveal the presence of these clusters in the low dimensional mapping.

t-SNE computations involve first computing similarity between observations in the high dimensional input space. The similarity of point $x_j$ to point $x_i$ is computed using the Gaussian conditional probability, $p_{j|i}$, in Equation (1), where $\sigma_i$ is the variance of the Gaussian distribution centered on $x_i$.

$$p_{j|i} = \frac{exp\left(-\|x_i - x_j\|^2/2\sigma_i^2\right)}{\sum_{k \neq i} exp(-\|x_i - x_k\|^2/2\sigma_i^2)} \quad (1)$$

In the next step, the similarity between two points $y_i$ and $y_j$ in low dimension is computed as the joint probabilities using the student-t distribution with one degree of freedom, as in Equation (2). Since the student-t distribution has thicker tails than the Gaussian distribution, it provides better discrimination for points that are farther spaced.

$$q_{ij} = \frac{\left(1 + \|y_i - y_j\|^2\right)^{-1}}{\sum_{k \neq l}(1 + \|y_k - y_l\|^2)^{-1}} \quad (2)$$

In the last step, the difference between similarities in original space and low-dimensional space is minimized by minimizing the Kullback-Liebler divergence (KL) to obtain co-ordinates of the points in lower dimensional space. For more details on mathematical formulation of t-SNE, refer to [5]. [6] provides an excellent tutorial on how to interpret t-SNE plots and describes effect of tuning parameters on results.

KPCA:

This section provides a brief overview of kernel principal component analysis (KPCA). The discussion in this section is based [7].
In the field of multivariate statistics, principal component analysis (PCA) is a widely used way to do linear dimensionality reduction. Its nonlinear counterpart, kernel principal component analysis (KPCA) [8] uses kernel methods and better exploits the complicated nonlinear structure of high-dimensional features. Both PCA and KPCA share the same underlying idea; that is, they all aim to project the set of data onto a low-dimensional subspace where the highest possible amount of data lies in. The major difference is that kernel PCA uses some mapping function to embed the data in a high-dimensional RKHS (reproducing kernel Hilbert space) called $\mathcal{F}$ and then implement linear dimension reduction with the "kernel trick" in $\mathcal{F}$. In doing this, we can find a nonlinear (with respect to the original input space) manifold where data is retained. The major application areas of KPCA include nonlinear dimensionality reduction, kernel principal component regression, image denoising, and novelty detection.

The ordinary KPCA involves eigen-decomposition with computational complexity $\mathcal{O}(n^3)$, where $n$ is the training data size. Therefore, for larger dataset KPCA can be very slow. To overcome this drawback, a method called Nyström low-rank approximation was proposed in [9]. It reduces computation complexity to $\mathcal{O}(nc^2)$, where $c$ is the sample size from original data. The sampling scheme is important to approximation result. In [10], a $k$-means sampling scheme is proposed and proves to be very effective. In most cases a small $c$ (around 100) is big enough to give a good approximation result. This low-rank approximation also reduces memory complexity of kernel PCA from $\mathcal{O}(n^2)$ to $\mathcal{O}(nc^2)$.

## 4 ANALYSIS

This section provides analysis of two data sets using kernel PCA and t-SNE. The first data set consists of vibration measurements taken on an experimental setup with a variable speed squirrel cage motor. The second data is the NASA turbo-fan degradation data set [13, 14].

### 4.1 Vibration Data Analysis

For this experiment, we collected data using a variable speed, three-phase Induction Squirrel Cage Motor equipped with three accelerometer vibration sensors owned by SAS (Figure 1). Using this machine, we ran three experiments with the motor running at 25, 35, and 50 revolutions per minute (rpm), respectively. While running each experiment, we collected data from the vibration sensors at 12,800 Hz (that is, 12,800 data reads per second). Initially we collected approximately 10 minutes of data with the machine running under its current, presumed non-fault, "normal" state. After approximately 10 minutes, we artificially changed the state of the motor from this normal condition. To do this, we adjusted the black knob on top of the blue plate near sensor Ai1. (Figure 2.)

This adjustment effectively changes the alignment of the main motor shaft ever so slightly from its "balanced" state.To understand what happens as this knob is adjusted, see Figure 2.

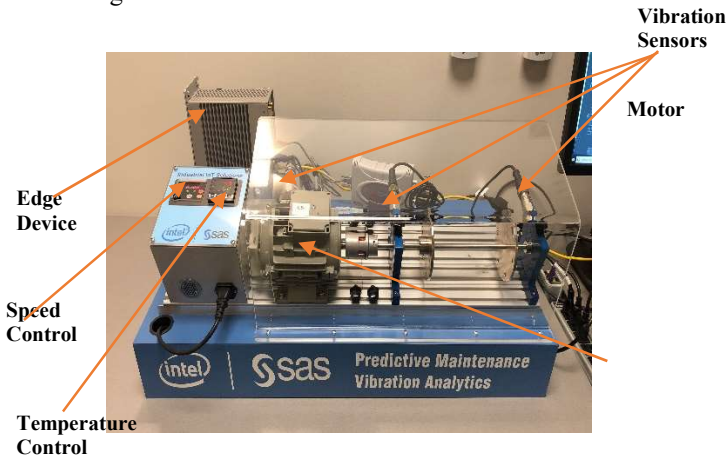

Figure 1. Squirrel Cage Motor System

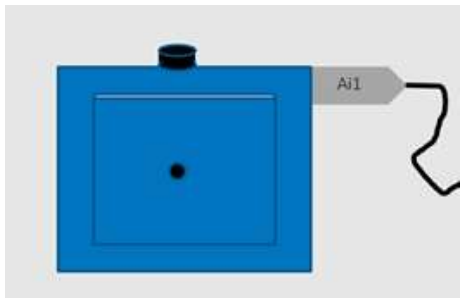

Figure 2. Knob to Adjust Shaft Balance

A twist of the knob raises the center rectangular section up slightly, changing the alignment of the shaft that runs through the hole illustrated in the center of this piece. In this way, we cause the shaft to become imbalanced. We then let the machine run in this new state with an artificially induced misalignment.

Once data were collected, we used a moving windows-based approach to analyze vibration data in the frequency domain using short-time Fourier transform (STFT). The window size was set to 4096 observations, and a new window was created by skipping 2048 observations. The window data were analyzed using STFT to get the frequency spectra. To reduce the noise in the spectra obtained, a moving average of the spectra obtained in 20 seconds was calculated. We then derived 13 features that characterize each averaged spectra by first dividing the range of frequencies covered by the spectra into 13 segments and then summing the magnitude (in dB) in each segment. The segmentation was performed using the SEGMENTATION function in the TSMODEL procedure of the SAS Time Series Analysis package on the "normal" cumulative magnitude curve from the 50 rpm experiment. Thirteen linear segments of varying length were obtained. These same frequency segments were used to derive the 13 model features for each averaged spectrum in all three speed experiments. For more details see [12].

### 4.1.2 Analysis

We analyzed the vibration data with 13 features, using t-SNE and KPCA. For visualization, we used 2-dimensional embedding for t-SNE and used loadings on the first two principal components obtained using the KPCA.

Figure 3 and Figure 4 illustrate the t-SNE results. We used the TSNE procedure in SAS Visual Data Mining and Machine Learning (VDMML) to analyze the data. Matten and Hinton in [5] state that "the performance of t-SNE is fairly robust to changes in perplexity, and typical values are between 5 and 50". We set the perplexity to 50. For learning rate, we used the default value of 100 available in the TSNE procedure. The data set used as an input to the t-SNE algorithm had 13 variables and 15,119 observations.

The plot in Figure 3 shows the 2-dimensional embedding obtained using the original 13 variables. The x and y axis values indicate the value of dimension one and two for each observation in the input data. The colors blue, red and green refer to machine rpm of 25, 35 and 50, respectively. The plot indicates that the 2-dimensional embedding obtained using t-SNE can identify the three disjoint clusters, each corresponding to a different rpm.

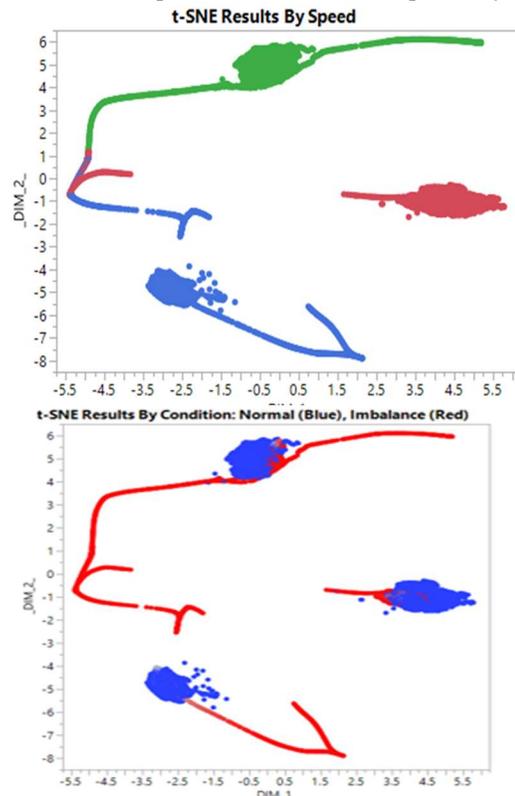

Figure 4. t-SNE results by Normal/Imbalance

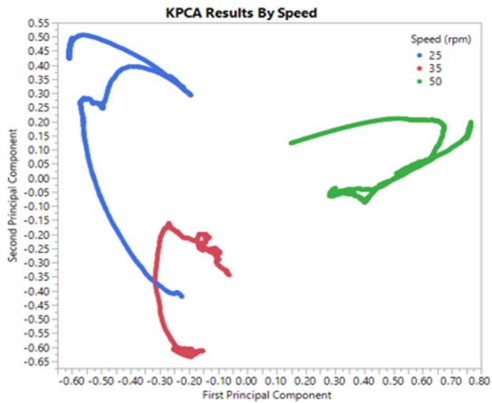

Figure 5: KPCA Results for Vibration Data

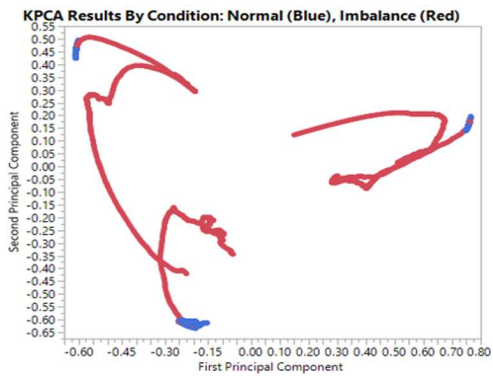

Figure 6: KPCA Results by Normal/Imbalance

Figure 4 shows the t-SNE results with respect to the normal conditions and the imbalanced condition induced as discussed in section 4.1. The "tails" emanating from the clusters refer to the "imbalanced" condition. The points in the "tails" that are closer to the clusters are observations immediately after the fault was induced. As the fault condition persisted, the observations moved along the tail. This is easily visible in an animated plot, which for obvious reasons we cannot include in this paper.

Figure 5 shows the plot obtained using the loading on the first two principal components of the KPCA analysis. Similar to the t-SNE, KPCA has identified three clusters corresponding to the three rpms.

Figure 6 shows the KPCA plot, where observations coming from the imbalanced condition are plotted in a separate color (red) than the observations coming from the normal conditions (blue color). Figure 6 shows that observations recorded when the shaft was imbalanced are on a trajectory (or on a "tail") away from the cluster of normal observations. To quantitatively evaluate the clustering quality of t-SNE and KPCA, we calculated the Davies–Bouldin index [11] for clusters of normal data. It is known that an algorithm produces clusters with low intra-cluster distances (high intra-cluster similarity) and high inter-cluster distances (low inter-cluster similarity) will have a low Davies–Bouldin index. Namely, the lower the index, the more compact and better seperated the clusters will be. For clusters of normal vibration data with different speeds, the t-SNE gives an index 0.1337 while KPCA gives an index 0.0215. Clearly, KPCA gives better clustering result in this example though both techniques give highly separated clusters.

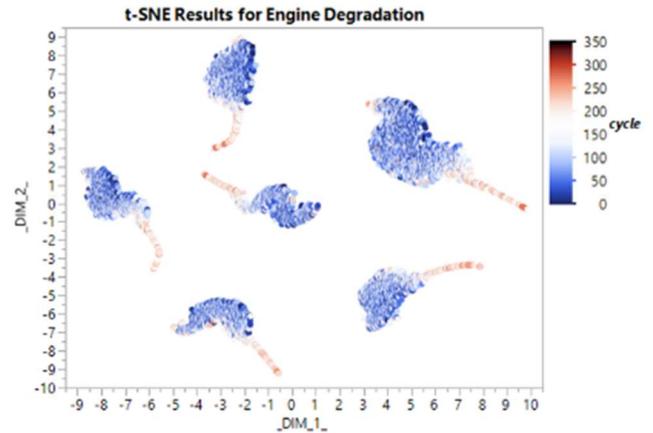

Figure 7: t-SNE Results for Engine Degradation Data

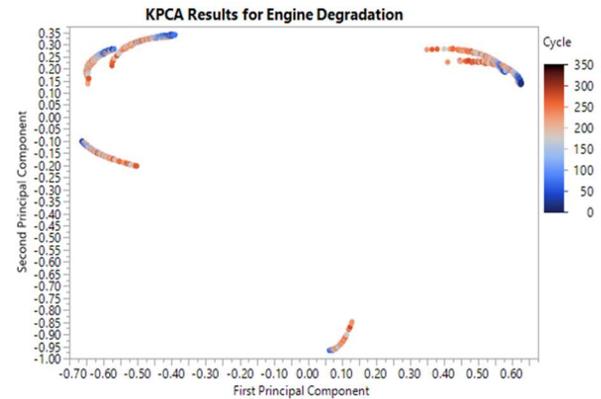

Figure 8: KPCA Results for Engine Degradation Data

### 4.2 *Analysis of Aircraft Engine Degradation Data*

In this section we analyze the aircraft engine degradation data from Saxena et al. (2008) [13] and Saxena and Goebel (2008) [14] using t-SNE and KPCA. The data that are used in this section consist of the flight history of 99 engines. Each flight is described by a vector of 24 variables. There are three variables related to the engine's operating conditions and 21 variables related to the sensor measurements. The dataset captures the entire history of each of the 99 aircraft engines, from the start of the service to the end of the service. The variable "cycle (flight)" is used to quantify the engine life. The number of flights until the end of the life is different for each engine since each engine

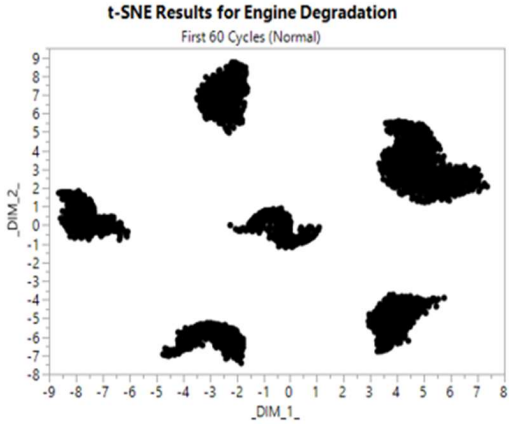

Figure 9: t-SNE Results for first 60 cycles

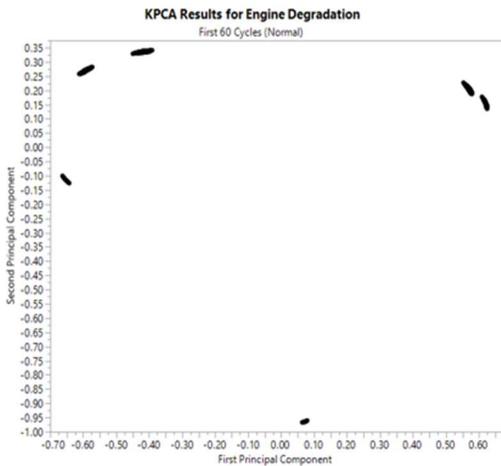

Figure 10: KPCA Results for first 60 cycles

degrades at a different rate.

### 4.2.2 Analysis

Similar to the vibration data, we analyzed this data with t-SNE and KPCA. For visualization, we used 2-dimensional embedding for t-SNE and used loadings on the first two principal components obtained using the KPCA.

Figure 7 illustrates the t-SNE result. A color scale is also provided to indicate the cycle sequence. We use the TSNE procedure in SAS Visual Data Mining and Machine Learning (VDMML) to analyze the data. We set the learning rate to 100 and perplexity to 5. The data set used as an input to the t-SNE algorithm has 24 variables and observations, corresponding to the engines with id 101 to 199. In Figure 7, the points in the tail represent the flights that were taken when engine was very near to its end of life. The six clusters found in the t-SNE plot correspond to six operating conditions of the engine.

Figure 8 illustrates the results using the loadings on the first two principal components obtained using the kernel PCA. In this plot we also find six clusters as in the t-SNE plot. The main difference between the two plots is that in the kernel PCA plot there is some overlapping in tails (degradation data) between different clusters.

We noticed that the shortest lived engine died at 139 cycles, thus here we take 60 cycles as a reasonable value when all machines are operating normally. We plotted the t-SNE and KPCA results of the first 60 cycles in Figure 9 and Figure 10, respectively. Figure 10 shows KPCA loadings of data belonging to normal cycles are still well separated and form six clusters.

Here again we calculated the Davies–Bouldin index for the normal part of engine degradation data (cycle<=60). KPCA gives an index 0.0933 and t-SNE gives 0.2554. However, if we put all the data together t-SNE will give a better clustering result since KPCA has overlap while t-SNE gives completely separated clusters, as shown in Figure 7 and Figure 8.

## 5 CONCLUSIONS

This paper illustrates the use of two machine learning techniques, namely the t-distributed stochastic neighbor embedding (t-SNE) and kernel principal component analysis (KPCA) for visualizing high dimensional sensor data vector. We demonstrated that both techniques were able to correctly identify the number of operating modes of the equipment.

When we analyzed data using these techniques, it was not required to specify the number of clusters in the data. Both techniques correctly identified the correct number of operating modes present in the input data. If the number of operating modes or disjoint clusters in the data is unknown during analysis, then these are powerful results.

Additionally, in each visualization the departure from the normal or stable state was captured in trajectories or "tails," and as equipment degraded more, the observations shifted farther away along these trajectories. Another important insight is that these visualizations can identify both the departure of sensor data from stable operations and the extent of departure. Both insights are important for fault detection and condition monitoring.

As noted earlier, both KPCA and t-SNE provide comparable results. However, KPCA provides certain advantages in practice. Most notably, with KPCA one can build a model using training data collected from normal conditions. The training data can then be projected on the first two principal components to get a low-dimensional visualization. As new data are collected from equipment, they can be easily scored and then plotted on the existing visualization. In comparison, t-SNE cannot be readily used to train and model and score new data efficiently. t-SNE involves constructing a probability distribution over pairs of high-dimensional objects and thus is fairly expensive computationally. The fast scoring available with KPCA makes it an ideal candidate for scoring streaming data on the edge in Internet of Things (IoT) applications.

KPCA has the added advantage that it provides much faster training when compared to t-SNE. The KPCA speed advantage is even greater when using the low-rank approximation outlined in section 3.

In summary, both approaches are useful for low-dimensional visualization of high dimensional multivariate sensor data. For ad-hoc analysis on low to medium volume data where scoring is not required, both approaches are equally effective. For real-time analysis of high volumes of sensor data, KPCA offers advantages with respect to fast training and scoring.

## REFERENCES

1. https://www.theglobeandmail.com/report-on-business/rob-magazine/the-future-is-smart/article24586994/
2. Chaudhuri, Arin, et al. "Sampling method for fast training of support vector data description." *2018 Annual Reliability and Maintainability Symposium (RAMS)*. IEEE, 2018.
3. Lee, Jong-Min, et al. "Nonlinear process monitoring using kernel principal component analysis." *Chemical engineering science* 59.1 (2004): 223-234.
4. Cho, Ji-Hoon, et al. "Fault identification for process monitoring using kernel principal component analysis." *Chemical engineering science* 60.1 (2005): 279-288.
5. Maaten, Laurens van der, and Geoffrey Hinton. "Visualizing data using t-SNE." *Journal of machine learning research* 9.Nov (2008): 2579-2605.
6. Wattenberg, Martin, Fernanda Viégas, and Ian Johnson. "How to use t-SNE effectively." *Distill* 1.10 (2016): e2.
7. Shen, Kai and Asgharzadeh. Zohreh, 2019. SAS Technical Paper: Kernel Principal Component Analysis Using SAS. Cary, NC: SAS Institute Inc.
8. Baker, C.T., 1977. The numerical treatment of integral equations.
9. Schölkopf, B., Smola, A. and Müller, K.R., 1998. Nonlinear component analysis as a kernel eigenvalue problem. *Neural computation*, 10(5), pp.1299-1319.
10. Zhang, K., Tsang, I.W. and Kwok, J.T., 2008, July. Improved Nyström low-rank approximation and error analysis. In *Proceedings of the 25th international conference on Machine learning* (pp. 1232-1239). ACM.
11. Davies, D. L., & Bouldin, D. W. (1979). A cluster separation measure. *IEEE transactions on pattern analysis and machine intelligence*, (2), 224-227.
12. McGuirk, A., Liao, Y., Biggs, B., Kakde, D., and Costin, J., 2019. Condition-Based Monitoring Using SAS® Event Stream Processing. In *SAS Conference Proceedings: SAS Global Forum 2019*.
13. Saxena, A., and Goebel, K. (2008). "Turbofan Engine Degradation Simulation Data Set." Accessed January 17, 2017. NASA Ames Prognostics Data Repository. NASA Ames Research Center, Moffett Field, CA https://ti.arc.nasa.gov/tech/dash/groups/pcoe/prognostic-data-repository/
14. Saxena, A., Goebel, K., Simon, D., and Eklund, N. (2008). "Damage Propagation Modeling for Aircraft Engine Run-to-Failure Simulation." In *Proceedings of the International Conference on Prognostics and Health Management, 2008*, 1–9. Piscataway, NJ: IEEE

## BIOGRAPHIES

Kai Shen
SAS Institute Inc.
100 SAS Campus Drive
Cary, North Carolina, 27513 USA

e-mail: Kai.Shen@sas.com

**Kai Shen** received his PhD degree in Statistics from North Carolina State University and now works as a Research Statistician Developer at SAS Institute.

Anya McGuirk
SAS Institute Inc.
100 SAS Campus Drive
Cary, North Carolina, 27513 USA

e-mail: Anya.McGuirk@sas.com

**Anya McGuirk** works as a Distinguished Research Statistician Developer at SAS Institute.

Yuwai Liao
SAS Institute Inc.
100 SAS Campus Drive
Cary, North Carolina, 27513 USA

e-mail: Yuwai.Liao@sas.com

**Yuwai Liao** works as a Senior Research Statistician Developer in the Internet of Things (IoT) department at SAS Institute.

Arin Chaudhuri
SAS Institute Inc.
100 SAS Campus Drive
Cary, North Carolina, 27513 USA

e-mail: Arin.Chaudhuri@sas.com

**Arin Chaudhuri** works as a Senior Manager at SAS Institute.

Deovrat Kakde
SAS Institute Inc.
100 SAS Campus Drive
Cary, North Carolina, 27513 USA

e-mail: Dev.Kakde@sas.com

**Deovrat Kakde** works as a Principal Research Statistician Developer at SAS Institute.